\documentclass[prd,aps,showpacs,preprintnumbers,amsmath, amssymb]{revtex4-2}

\usepackage{graphics,epsfig,hyperref}
\usepackage{graphicx}
\usepackage{dcolumn}
\usepackage{bm}
\usepackage{epstopdf}
\usepackage{color}
\usepackage{xcolor}
\usepackage{subfigure}
\usepackage{diagbox} 
\usepackage{adjustbox}
\usepackage{multirow}
\usepackage{amsmath,amssymb,amsfonts}

\usepackage[normalem]{ulem}
\usepackage{xcolor}

\usepackage{orcidlink}

\newcommand{\affHIAS}{School of Fundamental Physics and Mathematical Sciences,
Hangzhou Institute for Advanced Study, University of Chinese Academy of Sciences,
No.1 Xiangshan Branch, Hangzhou 310024, People's Republic of China}

\newcommand{\affHUNNU}{Department of Physics, Institute of Interdisciplinary Studies,
Key Laboratory of Low Dimensional Quantum Structures and Quantum Control of Ministry of Education,
Synergetic Innovation Center for Quantum Effects and Applications,
and Hunan Research Center of the Basic Discipline for Quantum Effects and Quantum Technologies,
Hunan Normal University, Changsha, Hunan 410081, People's Republic of China}

\newcommand{\affBNU}{School of Physics and Astronomy,
Beijing Normal University, Beijing 100875, China}

\newcommand{\affYZU}{Center for Gravitation and Cosmology,
College of Physical Science and Technology, Yangzhou University,
Yangzhou 225009, People's Republic of China}

\begin{document}
	\title{\boldmath Constraining Lorentz symmetry breaking in bumblebee gravity with extreme mass-ratio inspirals}
	
	\author{
        Sheng Long$^{1}$\,\orcidlink{0009-0009-0163-2724},
        Zhong-wu Xia$^{2}$\,\orcidlink{0009-0006-3764-1583},
        Huajie Gong$^{2}$\,\orcidlink{0000-0002-5903-9423},
        Zhoujian Cao$^{1,3,*}$\,\orcidlink{0000-0002-1932-7295},
        Qiyuan Pan$^{2,4,*}$,
        and Jiliang Jing$^{2,4,*}$\,\orcidlink{0000-0002-2803-7900}
    }

    \thanks{
        Corresponding authors:
        \href{mailto:zjcao@amt.ac.cn}{zjcao@amt.ac.cn},
        \href{mailto:panqiyuan@hunnu.edu.cn}{panqiyuan@hunnu.edu.cn},
        \href{mailto:jljing@hunnu.edu.cn}{jljing@hunnu.edu.cn}.
    }

    \affiliation{$^{1}$\affHIAS}
    \affiliation{$^{2}$\affHUNNU}
    \affiliation{$^{3}$\affBNU}
    \affiliation{$^{4}$\affYZU}

	\begin{abstract}
	\baselineskip=0.6 cm
	\begin{center}
		{\bf Abstract}
	\end{center}	
     
Extreme mass-ratio inspirals (EMRIs), with their long-lived and highly relativistic orbital evolution, can probe strong-field spacetime geometry and provide an important means to test general relativity. In this work, we investigate EMRI waveforms in a Schwarzschild-like black hole spacetime arising in bumblebee gravity, where Lorentz symmetry breaking (LSB) is characterized by a dimensionless parameter $\ell$. We construct EMRI waveforms within the Augmented Analytic Kludge (AAK) framework using the modified orbital frequencies and fluxes. We find that  $\ell$ significantly affects the orbital  evolution and thereby modifies the waveform. These modifications grow with increasing $\ell$ and are further enhanced for more eccentric orbits. Furthermore, using Bayesian analysis, we obtain the posterior distributions of EMRI with the parameter $\ell$ included. Our results show that all injected source parameters are recovered within their $1\,\sigma$ credible intervals. We find that the bumblebee parameter $\ell$ can be constrained with an uncertainty of order {$\mathcal{O}(10^{-4})$} by LISA.

     \end{abstract}

\maketitle

\flushbottom

\section{Introduction}

In 2015, LIGO made the first direct detection of gravitational waves (GWs) from the binary black hole merger event GW150914~\cite{LIGOScientific:2016aoc}. Since then, the LIGO–Virgo–KAGRA (LVK) Collaboration has reported a rapidly growing catalog of GW events~\cite{Wadekar:2023gea,Mehta:2023zlk,Olsen:2022pin,Nitz:2021zwj,KAGRA:2021vkt,LIGOScientific:2017vwq,LIGOScientific:2014pky,VIRGO:2014yos,KAGRA:2020tym}. Next-generation ground-based detectors, such as the Einstein Telescope~\cite{2010CQGra..27s4002P} and Cosmic Explorer~\cite{Evans:2021gyd}, are expected to further increase the detection rate and extend the observable distance of compact binaries~\cite{Kalogera:2021bya,Reitze:2019iox}.
 Meanwhile, space-based detectors, including LISA~\cite{LISAConsortiumWaveformWorkingGroup:2023arg}, TianQin~\cite{TianQin:2015yph}, and Taiji~\cite{TaijiScientific:2021qgx}, will observe GWs in the millihertz band and enable the detection of sources including extreme mass-ratio inspirals (EMRIs).
The EMRI is a system in which a stellar-mass compact object ($1-10^2\,M_\odot$) gradually inspirals into a massive black hole ($10^4-10^7\,M_\odot$) due to GW emission. Since the smaller object spends a long time in the strong-field region, the system can undergo $10^{4}-10^{5}$ orbital cycles over several years of observation, accumulating a large amount of phase information in the waveform~\cite{Finn:2000sy,Berry:2016bit,Ye:2023lok,Fu:2024cfk, Dotti:2012qw,Barack:2018yly}. As a result, EMRI signals are highly sensitive to the spacetime geometry near the central black hole, making them effective in probing strong-field gravity and suited for testing general relativity (GR) and its potential deviations~\cite{Amaro-Seoane:2019umn,Babak:2017tow,Vazquez-Aceves:2022dgi,Zi:2025jxy,Zou:2025fsg,Zi:2025onl,Liu:2025swi,Meng:2025glf,Zhao:2025ktx,Fan:2025ymm,Guo:2022euk,Liang:2025ovq,Gair:2012nm,Heidari:2026eim,Meng:2025glf,Meng:2024kug,Zhang:2026ekj}. 

Although GR has achieved remarkable success in explaining current observations, it remains theoretically incomplete, owing to critical issues such as its
incompatibility with quantum theory and the unresolved nature of the dark sector of the Universe~\cite{Will:2014kxa}. These issues motivate the exploration of modified theories of gravity~\cite{Sotiriou:2008rp}. Among various possibilities, Lorentz symmetry breaking (LSB) provides a well-motivated framework for probing possible low-energy remnants of quantum gravity. Although Lorentz symmetry is a cornerstone of both the Standard Model and GR, several approaches to quantum gravity suggest that it may be spontaneously or dynamically broken near the Planck scale~\cite{Amelino-Camelia:2008aez,PhysRevD.39.683,Liberati:2013xla,Mattingly:2005re}. LSB can be systematically described within the Standard-Model Extension(SME), proposed by Kostelecký and collaborators~\cite{Kostelecky:2003fs,Colladay:1998fq}. In the gravitational sector of the SME, a simple and widely studied realization is the bumblebee gravity model \cite{Liu:2022dcn,Liu:2024wpa,Liu:2025bpp,Liu:2024oeq,Deng:2025uvp,Filho:2022yrk,AraujoFilho:2024ykw,AraujoFilho:2025rvn,AraujoFilho:2025zaj,Shi:2025plr}, in which a vector field acquires a nonzero vacuum expectation value and thereby spontaneously breaks local Lorentz symmetry~\cite{PhysRevD.39.683,PhysRevD.40.1886,Bluhm:2004ep}. This mechanism modifies the gravitational field equations and gives rise to a black hole solution which differs from the Schwarzschild case. 
Although such solutions have been investigated through orbital dynamics and weak-field tests~\cite{QiQi:2026zwp}, we instead use EMRI waveforms to constrain LSB effects in bumblebee gravity.

To investigate the capability of EMRIs to probe bumblebee gravity, efficient waveform models are required for data analysis. Realistic EMRI detection requires waveform templates that incorporate the second-order self-force, but constructing such waveforms in Schwarzschild spacetime remains a major challenge~\cite{Poisson:2004gg,Wald:2009ue,Barack:2018yvs,Poisson:2011nh,Wei:2025lva,Zhang:2026gdk}. In the absence of fully self-consistent relativistic waveforms in bumblebee gravity, semi-analytic waveform models offer a practical alternative by balancing computational efficiency and physical accuracy. In particular, the Augmented Analytic Kludge (AAK) model has been widely adopted for EMRI waveform generation and data analysis~\cite{Babak:2006uv,Barack:2003fp,Chua:2015mua,Chua:2016jnd, Zi:2025jxy,Zi:2023omh, Zi:2022hcc, Zhang:2026lrd, Fu:2024cfk, Zhang:2024ogc, Zhang:2022rfr}. The full implementation is publicly available through the \texttt{FastEMRIWaveforms}~\cite{Katz:2021yft,Chua:2020stf,Speri:2023jte,Chapman-Bird:2025xtd,BHPToolkit}.

Specifically, we first derive the conservative orbital motion of a test body in the bumblebee-modified spacetime and then compute the adiabatic inspiral driven by the corresponding energy and angular momentum fluxes. These ingredients are incorporated into the AAK framework to generate EMRI waveforms, in which the bumblebee parameter enters through the modified radial frequency and the evolution of the orbital parameters. The resulting waveforms are then projected onto the detector response and used as templates for the subsequent parameter estimation. {For parameter estimation, we use Markov Chain Monte Carlo (MCMC) sampling in a reduced five-dimensional intrinsic-parameter space rather than a Fisher-matrix approximation~\cite{Kejriwal:2023djc,Speri:2024qak,Fan:2025ymm,Xia:2026aty}. }

In this work, we investigate EMRI waveforms in a bumblebee black hole spacetime and develop an analysis pipeline from orbital dynamics to parameter estimation. In Sec.~\ref{sec:LQGBH}, we derive the orbital dynamics and post-Newtonian (PN) expansions of the relevant frequencies and phases. In Sec.~\ref{sec:fluxes_evolution}, we compute the energy and angular momentum fluxes and obtain the adiabatic evolution equations for the orbital parameters. In Sec.~\ref{sec:waveform}, we construct EMRI waveforms within the AAK framework. In Sec.~\ref{sec:WA}, we show how the bumblebee parameter $\ell$ enters the waveform  through the orbital dynamics and phase evolution. In Sec.~\ref{sec:WB}, we first quantify waveform differences in terms of the mismatch, and then use a Bayesian analysis to constrain the detectability of intrinsic parameters. In Sec.~\ref{sec:con} we summarize our results.

\section{Spacetime and orbital dynamics in bumblebee gravity}
\label{sec:LQGBH}

We consider a Schwarzschild-like black hole solution in bumblebee gravity~\cite{Colladay:1998fq,Kostelecky:2003fs}. 
In this framework, a vector field acquires a nonzero vacuum expectation value, 
which modifies the gravitational field equations. We adopt the static, spherically symmetric solution obtained in Ref.~\cite{Casana:2017jkc}, 
corresponding to a radial bumblebee background. The resulting metric takes the form
\begin{equation}
	ds^2 = - f(r)\,dt^2 + \frac{1+\ell}{f(r)}\,dr^2 
	+ r^2\left(d\theta^2 + \sin^2\theta\,d\varphi^2\right),
	\label{eq:metric}
\end{equation}
where
\begin{equation}
	f(r) = 1 - \frac{2M}{r},
\end{equation}
and $M$ is the black hole mass. The dimensionless parameter $\ell$ 
originates from the nonzero vacuum expectation value of the bumblebee field, 
{with $\ell>-1$. The interval $\ell\in[-0.5,0.5]$ adopted below is a computational prior, not an observationally motivated range. For the same static solution, Ref.~\cite{Casana:2017jkc} reported an attainable Cassini sensitivity of $\ell<5.9\times10^{-13}$. A full 14-dimensional analysis of the S2 orbit reported best-fit values of order $10^{-5}$, with $1\sigma$ uncertainties of approximately $2\times10^{-4}$~\cite{QiQi:2026zwp}. Comparing these results requires the assumption that the same parameter $\ell$ applies in weak- and strong-field systems. With that assumption, the Cassini sensitivity is substantially tighter, while the EMRI calculation probes a different gravitational regime.}

The dynamics of a test particle with mass $m$ in the spacetime~\eqref{eq:metric} is governed by the Lagrangian~\cite{chand1998mtbh}
\begin{equation}
	\mathcal{L} = \frac{1}{2} m\, g_{\mu\nu}\dot{x}^\mu\dot{x}^\nu,
	\label{eq:lagrangian}
\end{equation}
where an overdot denotes differentiation with respect to the proper time $\tau$ along timelike geodesics. Owing to the time-translation and rotational symmetries, the orbital energy $E$ and axial angular momentum $L_z$ are conserved quantities associated with the Killing vectors $\partial_t$ and $\partial_\varphi$, respectively. Restricting the motion to the equatorial plane $\theta = \pi/2$, the geodesic equations give the radial equation of motion
\begin{equation}
	\dot{r}^2 =\frac 1{(1+\ell)}\left[ \frac{E^2}{m^2}
	- f(r)\left(1 + \frac{L_z^2}{m^2 r^2}\right)\right],
	\label{eq:radial_EOM}
\end{equation}
together with
\begin{equation}
	\dot{t} = \frac{E}{m f(r)},\qquad
	\dot{\varphi} = \frac{L_z}{m r^2}.
	\label{eq:tphi_EOM}
\end{equation}

For bound orbits, we introduce the semilatus rectum $p$, eccentricity $e$, and parameterize the radius in terms of $\chi$~\cite{Babak:2006uv}
\begin{equation}
	r(\chi) = \frac{p\,M}{1 + e\cos\chi}.
	\label{eq:rel_anomaly}
\end{equation}
{Throughout this paper, $p$ denotes the dimensionless semi-latus rectum; the corresponding length scale is $pM$.}
The periapsis and apoapsis are
\begin{equation}
	r_{\rm p} = \frac{p\,M}{1 + e},\qquad
	r_{\rm a} = \frac{p\,M}{1 - e}.
	\label{eq:rp_ra}
\end{equation}
The conserved quantities $E$ and $L_z$ follow from imposing $\dot r(r_{\rm p}) = \dot r(r_{\rm a}) = 0$ in Eq.~\eqref{eq:radial_EOM}, yielding
\begin{align}
	\frac{E^2}{m^2} &= \frac{(p - 2 - 2e)(p - 2 + 2e)}{p\left(p - 3 - e^2\right)}, \\
	\frac{L_z^2}{m^2} &= \frac{M^2 p^2}{p - 3 - e^2}.
\end{align}
These expressions have the same form as in the Schwarzschild case.

The radial and azimuthal orbital frequencies are defined by
\begin{equation}
	\Omega_r=\frac{2\pi}{T_r},\qquad
	\Omega_\varphi=\frac{\Delta\varphi}{T_r},
	\label{eq:freq_def}
\end{equation}
where \(T_r\) denotes the radial period and \(\Delta\varphi\) is the azimuthal phase accumulated during one radial cycle. In terms of  \(\chi\), they are given by
\begin{equation}
	T_r=2\int_{r_{\rm p}}^{r_{\rm a}}\frac{dt}{dr}\frac{dr}{d\chi}\,d\chi,\qquad
	\Delta\varphi=2\int_{r_{\rm p}}^{r_{\rm a}}
	\frac{d\varphi}{dt}\frac{dt}{dr}\frac{dr}{d\chi}\,d\chi.
	\label{eq:Tr_Dphi}
\end{equation}
These quantities admit weak-field expansions in powers of $p$:
\begin{align}	
	\frac{T_r}{\sqrt{1+\ell}} &=\frac{2\pi M}{X^3} p^{3/2}
	+\frac{6\pi M}{X} p^{1/2}
	+\frac{3\pi M(5+4/X)}{p^{1/2}}\notag+\frac{\pi M\left[3(16+25X+4X^2)\right]}{p^{3/2} X}
	+ \mathcal{O}(p^{-5/2}),\\
	 \Omega_r\sqrt{1+\ell}&= \frac{X^{3}}{Mp^{3/2}}
	-\frac{3X^{5}}{Mp^{5/2}}
	+\frac{3X^{5}(-4-5X+6X^{2})}{2Mp^{7/2}}\notag\\&+\frac{-3X^{5}(16+25X-20X^{2}-30X^{3}+18X^{4})
		}{2Mp^{9/2}}
	+ \mathcal{O}(p^{-11/2}), \\[0.4em]
	\Omega_\varphi &= \frac{X^{3}}{Mp^{3/2}}
	+\frac{-6X^{3}(-1+X^{2})}{2Mp^{5/2}}+\frac{6X^{3}(19-21X^{2}-10X^{3}+12X^{4})
		}{8Mp^{7/2}}
	+ \mathcal{O}(p^{-9/2}).
\end{align}
where $X=\sqrt{1-e^2}$. Within the AAK framework, it is necessary to incorporate the corrections to the orbital frequencies induced by the bumblebee parameter. Interestingly, the above results indicate that the bumblebee parameter leaves the azimuthal frequency \(\Omega_\varphi\) unchanged, while the radial frequency receives only an overall correction in the form of a constant rescaling factor \(1/\sqrt{1+\ell}\).

\section{Gravitational-wave fluxes and orbital evolution}
\label{sec:fluxes_evolution}

GW emission causes the orbit to shrink and gradually circularize over time. Within the adiabatic approximation~\cite{isoyama2022adiabatic}, the inspiral can be described as a sequence of geodesics, with the orbital parameters $(p,e)$ evolving slowly due to the loss of energy and angular momentum.

At leading order, the energy and angular-momentum fluxes are computed using the quadrupole formula~\cite{thorne1980multipole,maggiore2008gravitational},
\begin{equation}
	-\frac{dE}{dt} = \frac{1}{5} \left\langle \dddot{Q}_{ij} \dddot{Q}^{ij} \right\rangle,
	\quad -\frac{dL_i}{dt} = \frac{2}{5} \epsilon_{ijk} \left\langle \ddot{Q}^{jm} \dddot{Q}^k_{\quad\!\!\!\! m} \right\rangle,\label{eq:flux}
\end{equation}
using the trajectory $x^i(t) = (r\cos\varphi,r\sin\varphi,0)$, where the indices $i$, $j$, $k$, and $m$ are summed over spatial components only. Here $Q_{ij} = m \left( x_i x_j - \frac{1}{3} \delta_{ij} x^kx_k \right)$ is the mass quadrupole moment, $\epsilon_{ijk}$ is the Levi-Civita symbol, and $\langle \cdot \rangle$ denotes averaging over one orbital period.

After time-averaging, we can express the fluxes as expansions in powers of $1/p$ and the eccentricity. The bumblebee correction enters the fluxes through the modified orbital trajectory used in the quadrupole radiation formula.
{This treatment is phenomenological. We evaluate the standard quadrupole formula on the modified orbit but do not derive the dissipative dynamics from the coupled metric--bumblebee perturbation equations. The model therefore omits bumblebee-field radiation, additional polarization modes, propagation corrections, and a theory-specific self-force. The forecast assumes that the modified orbital dynamics provides the leading observable deviation.}
{Within this approximation, we obtain}
\begin{align}\label{eq:energy_flux}
	\left\langle \frac{dE}{dt} \right\rangle =& \frac {(1 - e^2)^{3/2}m^2}{15p^5M^2(1+\ell)^3}\left[96 + 292 e^2 + 37 e^4 
	+ 12 \left(24 + 104 e^2 + 33 e^4 \right) \ell 
	+ 6 \left(48 + 280 e^2 + 150 e^4 + 5 e^6 \right) \ell^2 \right. \notag\\
	&\left. + 6 \left(16 + 120 e^2 + 90 e^4 + 5 e^6 \right) \ell^3
	\right]+ \frac{(1 - e^2)^{3/2}m^2}{5p^6M^2} \left[e^2 \left(176 + 450 e^2 + 53 e^4\right) \right. \notag\\
	&\left.
	+ e^2 \left(-216 + 732 e^2 + 437 e^4\right)\ell
	+ 2 e^2 \left(-592 - 600 e^2 + 190 e^4 + 15 e^6\right)\ell^2\right. \notag\\
	&\left.
	+ 6 e^2 \left(-128 - 240 e^2 + 5 e^6\right)\ell^3\right]
	+ \mathcal{O}(p^{-7}),\\
	\left\langle \frac{dL_z}{dt} \right\rangle =& \frac{4(1 - e^2)^{3/2}m^2}{5 p^{7/2}M(1+\ell)^2}\left[8 + 7 e^2 
	+ \left(16 + 28 e^2 + e^4\right)\ell 
	+ \left(8 + 24 e^2 + 3 e^4\right)\ell^2\right]+ \frac{4e^2(1 - e^2)^{3/2} m^2}{5 p^{9/2}M(1+\ell)^2} \notag\\
	&\times \left[38 + 27 e^2 
	+ \left(-16 + 66 e^2 + 3 e^4\right)\ell 
	+ \left(-88 - 12 e^2 + 9 e^4\right)\ell^2
	\right]
	+ \mathcal{O}(p^{-11/2}).\label{eq:angular_flux}
\end{align}
 For brevity, we display only the first leading two PN orders here, while higher-order terms are also included in the numerical implementation.

The orbital evolution equations for $(p,e)$ follow from
\begin{align}
	\frac{dE}{dt} &= \frac{\partial E}{\partial p}\frac{dp}{dt}
	+ \frac{\partial E}{\partial e}\frac{de}{dt}, \label{eq:dEdt}
    \\[0.3em]
	\frac{dL_z}{dt} &= \frac{\partial L_z}{\partial p}\frac{dp}{dt}
	+ \frac{\partial L_z}{\partial e}\frac{de}{dt},
	\label{eq:dEdt_dLdt}
\end{align}
Solving Eqs.~\ref{eq:dEdt} and \ref{eq:dEdt_dLdt} for $dp/dt$ and $de/dt$, we obtain the adiabatic evolution of orbital parameters
\begin{align}
	\left\langle \frac{dp}{dt}  \right\rangle=&
	- \frac{8 \left(1 - e^2\right)^{3/2}  m}{5  p^3M(1+\ell)^2}\left[8 + 7 e^2 
	+ \left(16 + 28 e^2 + e^4\right)\ell 
	+ \left(8 + 24 e^2 + 3 e^4\right)\ell^2
	\right]-\frac{2 \left(1 - e^2\right)^{3/2} m}{15   p^4M(1+\ell)^3} \notag \\
	&\times \left[144 + 326 e^2 + 245 e^4
	+ 6 \left(72 - 13 e^2 + 90 e^4 + 5 e^6\right)\ell
	+ 6 \left(72 - 252 e^2 - 74 e^4 + 15 e^6\right)\ell^2\right. \notag\\
	&\left.
	+ 6 \left(24 - 184 e^2 - 123 e^4 + 10 e^6\right)\ell^3
	\right]+ \mathcal{O}(p^{-5}),\label{eq:dpdt}\\
	\left\langle \frac{de}{dt}  \right\rangle  =& -\frac{e(1-e^2)^{3/2}m}{15p^4M(1+\ell)^3}
	\left[304 + 121 e^2
	+ 12 \left(93 + 67 e^2 + e^4\right)\ell
	+ 6 \left(224 + 246 e^2 + 13 e^4\right)\ell^2\right. \notag\\
	&\left.
	+ 66 \left(8 + 12 e^2 + e^4\right)\ell^3\right]
	-\frac{e(1-e^2)^{3/2}m}{30 p^5M(1+\ell)^3} \left[1280 + 2097 e^2 + 697 e^4
	+ 6 \left(434 + 100 e^2 + 409 e^4 + 10 e^6\right)\ell\right. \notag\\
	&\left.
	+ 6 \left(160 - 1946 e^2 - 233 e^4 + 45 e^6\right)\ell^2
	+ 6 \left(-40 - 1652 e^2 - 521 e^4 + 35 e^6\right)\ell^3
	\right]+ \mathcal{O}(p^{-6})\label{eq:dedt}.
\end{align}
The leading-order term reproduces the standard GR inspiral, while the $\ell$-dependent terms modify the rate of orbital decay and eccentricity reduction.

\section{EMRI waveforms in the bumblebee-modified spacetime}
\label{sec:waveform}

\subsection{Augmented Analytic Kludge Waveform}
\label{sec:WA}

A variety of EMRI waveform models have been developed {over the past decades}, including the Analytic Kludge (AK)~\cite{peters1963gravitational,peters1964gravitational}, Numerical Kludge (NK)~\cite{Babak:2006uv}, and {Teukolsky waveform}~\cite{hughes2021adiabatic,isoyama2022adiabatic,nasipak2024adiabatic}, with continuous efforts aimed at improving accuracy and computational efficiency~\cite{khalvati2025impact}. Among these, the AAK model~\cite{chua2017augmented} is particularly effective: by incorporating {frequency corrections derived from the NK model} into the AK framework, {it achieves relatively high accuracy at low computational cost, making it well suited for large-scale LISA template generation}.

Although fully relativistic self-force waveforms provide the highest precision~\cite{Katz:2021yft,Chapman-Bird:2025xtd}, {their computation is extremely challenging.} 
{As a result, self-force studies in modified gravity are still at an early stage~\cite{Roy:2025kra}}, motivating the continued use of PN-based kludge expansions. These approximations capture the essential strong-field dynamics relevant for EMRI systems. 
{An advantage of the AAK model is that it allows the rapid generation of a large number of templates, making it suitable for Bayesian analyses of the constraining capability of EMRI observations on bumblebee gravity.}

Following Refs.~\cite{Babak:2006uv,Sopuerta:2009iy}, we use the multipolar expansion of metric perturbations and retain only the leading {mass quadrupole} term. The transverse–traceless metric perturbation is then
\begin{equation}
	h_{i j}=\frac{2}{D}\left(P_{i k} P_{j l}-\frac{1}{2} P_{i j} P_{k l}\right) \ddot{I}^{k l}
\end{equation}
where $I_{ij}$ is the mass quadrupole, $D$ is the luminosity distance and $P_{i j}$ is the projection operator defined as $P_{i j}=\delta_{i j}-\hat{n}_i \hat{n}_j$. For a point particle, the quadrupole moment is
\begin{equation}
	I^{ij}=\left[\int d^3x\, x^i x^j\, T^{tt}(t,x^k)\right]^{\mathrm{STF}},
\end{equation}
with the {energy density} given by
\begin{equation}
	T^{tt}(t,x^i)=m \,\delta^{(3)}\!\left[x^i-z^i(t)\right],
\end{equation}
{where $z^i(t)$ is the  position of particle}.

{The NK approach treats the motion as a “bead on a wire,” using a curved-spacetime trajectory as input to a flat-space wave-generation formula. Despite its formal inconsistency, it is generally quite accurate, with overlaps exceeding $95\%$ against Teukolsky-based waveforms over a significant part of the relevant parameter space~\cite{Babak:2006uv}. } 
{While the NK provides a much more faithful approximation to EMRI waveforms, its construction is still significantly more expensive than that of the analytic kludge. This motivates the AAK, which incorporates key information from NK to improve waveform faithfulness without sacrificing the speed that makes kludge models attractive for data analysis.}
Then, in AAK the GW signal is written as a sum over harmonics of the radial and azimuthal motions~\cite{chua2017augmented}
\begin{equation}
	h_{+,\times}(t) = \sum_{n=1}^{\infty} A^{+,\times}_{n}(t),
	\label{eq:h_decomposition}
\end{equation}
{In the numerical implementation, we follow the default AAK prescription $n_{\max}=\lfloor 30e_0\rfloor$~\cite{chua2017augmented}. For the largest initial eccentricity considered here, $e_0=0.2$, this prescription gives $n_{\max}=6$, which we adopt for the waveform generation, mismatch calculation, and MCMC analysis. Appendix~\ref{app:nmax_convergence} presents the convergence tests for this choice.}
The harmonic amplitudes $A_n^{+}$ and $A_n^{\times}$ are given by
\begin{align}
	A_n^+ &= \left[ 1 + (\hat{L} \cdot \hat{n})^2 \right] \left[ b_n \sin(2\gamma) - a_n \cos(2\gamma) \right] + \left[ 1 - (\hat{L} \cdot \hat{n})^2 \right] c_n, \label{eq:hplus}\\
	A_n^\times &= 2 (\hat{L} \cdot \hat{n}) \left[ b_n \cos(2\gamma) + a_n \sin(2\gamma) \right],\label{eq:hcross}
\end{align}
where
\begin{align}
	&\gamma=\varphi_\varphi(t)-\varphi_r(t),\qquad
	\varphi_r(t)=\int_0^t \Omega_r(t') \, d t',\qquad
	\varphi_\varphi(t)=\int_0^t \Omega_\varphi(t') \, dt',\\
	&\hat{L} \cdot \hat{n}
	=\cos\theta_K\cos\theta_S+\sin\theta_K\sin\theta_S\cos(\varphi_K-\varphi_S).\notag
\end{align}
	
{We fix the sky position and orbital-angular-momentum direction to $\theta_S=\pi/3$, $\varphi_S=\pi/2$, $\theta_K=\pi/4$, and $\varphi_K=\pi/4$. The phase definitions above imply $\varphi_r(0)=\varphi_\varphi(0)=0$. These quantities and the luminosity distance $D=1\,\mathrm{Gpc}$ are not sampled in the inference.}
For eccentric motion, the coefficients take the form
\begin{align}
	a_n &= -n \mathcal A \left[ J_{n-2}(ne) - 2e J_{n-1}(ne) + 2 J_n(ne)/n + 2e J_{n+1}(ne) - J_{n+2}(ne) \right] \cos(n\varphi_r(t)), \notag\\
	b_n &= -n \mathcal A \sqrt{1 - e^2} \left[ J_{n-2}(ne) - 2J_n(ne) + J_{n+2}(ne) \right] \sin(n\varphi_r(t)), \\
	c_n &= 2 \mathcal A J_n(ne) \cos(n\varphi_r(t)),\notag
\end{align}
where $J_n(ne)$ is a Bessel function of the first kind, the amplitude $\mathcal A=(M\Omega_\varphi)^{2/3}m/D$ and luminosity distance $D=1 \,\mathrm{Gpc}$.

{The radial frequency $\Omega_r$ and the orbital parameters $(p,e)$ are directly modified by the bumblebee gravity, while the azimuthal frequency $\Omega_\varphi$ retains unchanged.}

\begin{figure}[tbp]
	\centering
	\begin{minipage}[b]{0.48\textwidth}
		\centering
		\includegraphics[width=\textwidth]{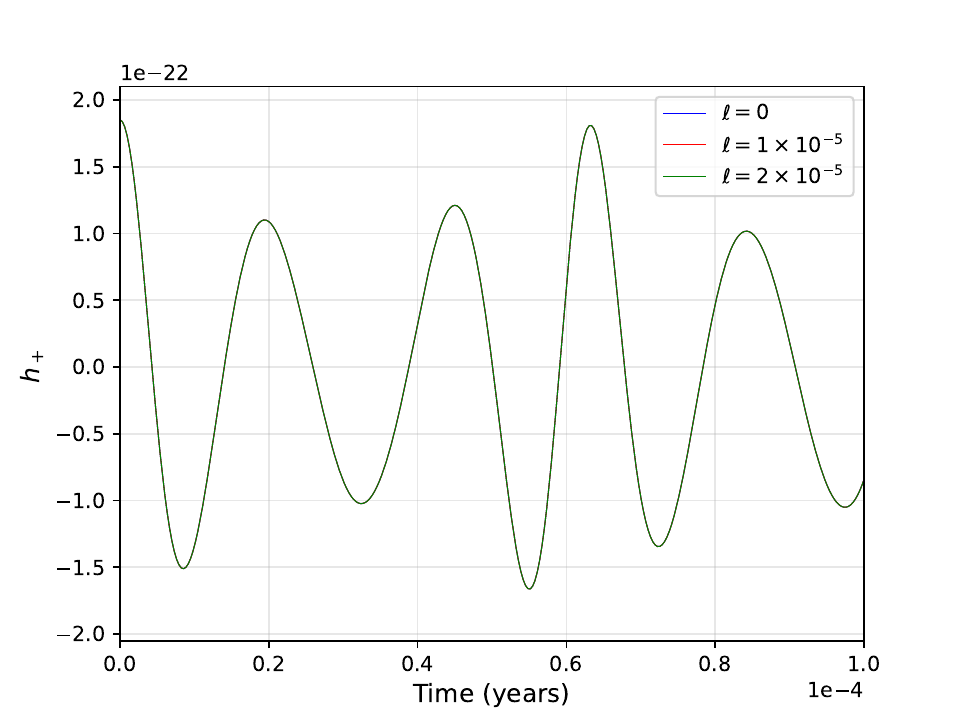}
	\end{minipage}
	\hfill
	\begin{minipage}[b]{0.48\textwidth}
		\centering
		\includegraphics[width=\textwidth,trim=0 19 0 0,clip]{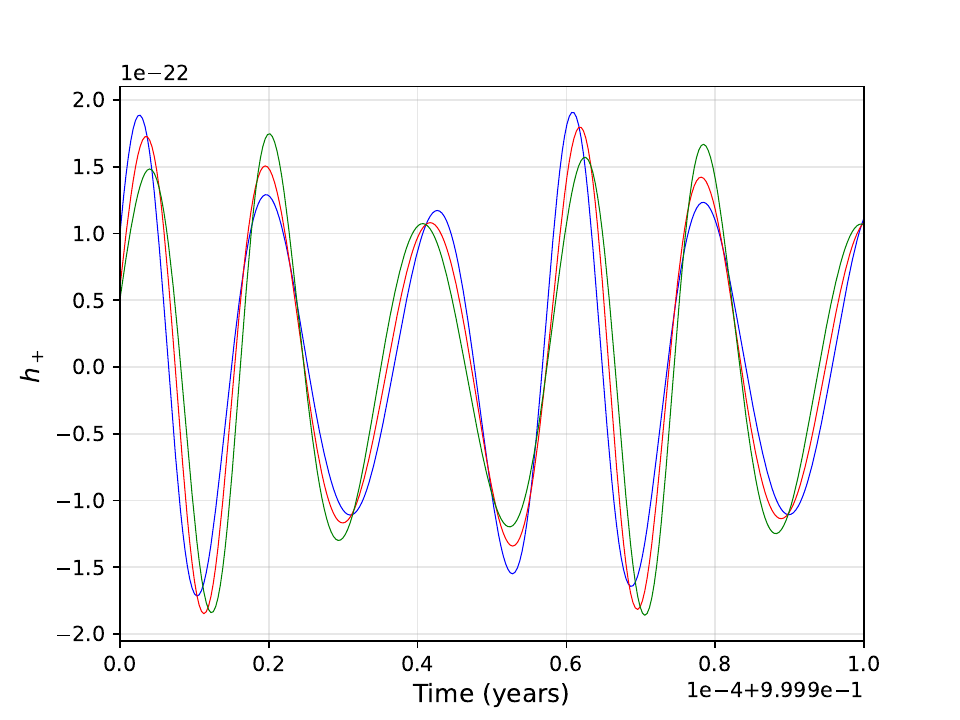}\\[-0.2em]
		\makebox[\textwidth]{\scriptsize $(t-t_0)/(10^{-4}\,{\rm yr})$}
	\end{minipage}
	\vspace{0.5em}
	\begin{minipage}[b]{0.48\textwidth}
		\centering
		\includegraphics[width=\textwidth]{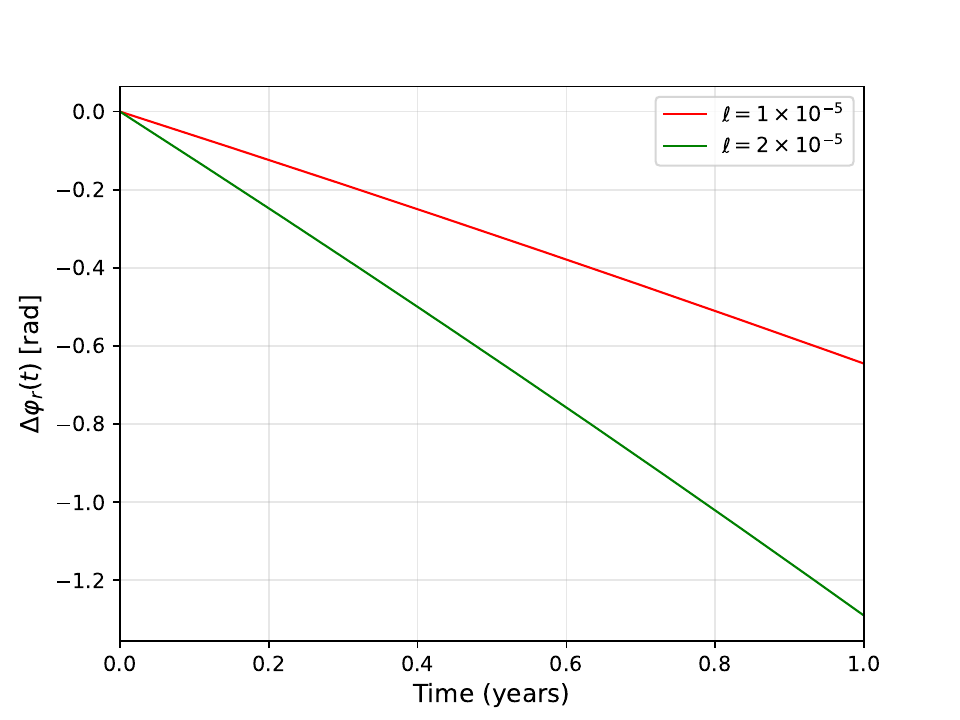}
	\end{minipage}
	\hfill
	\begin{minipage}[b]{0.48\textwidth}
		\centering
		\includegraphics[width=\textwidth]{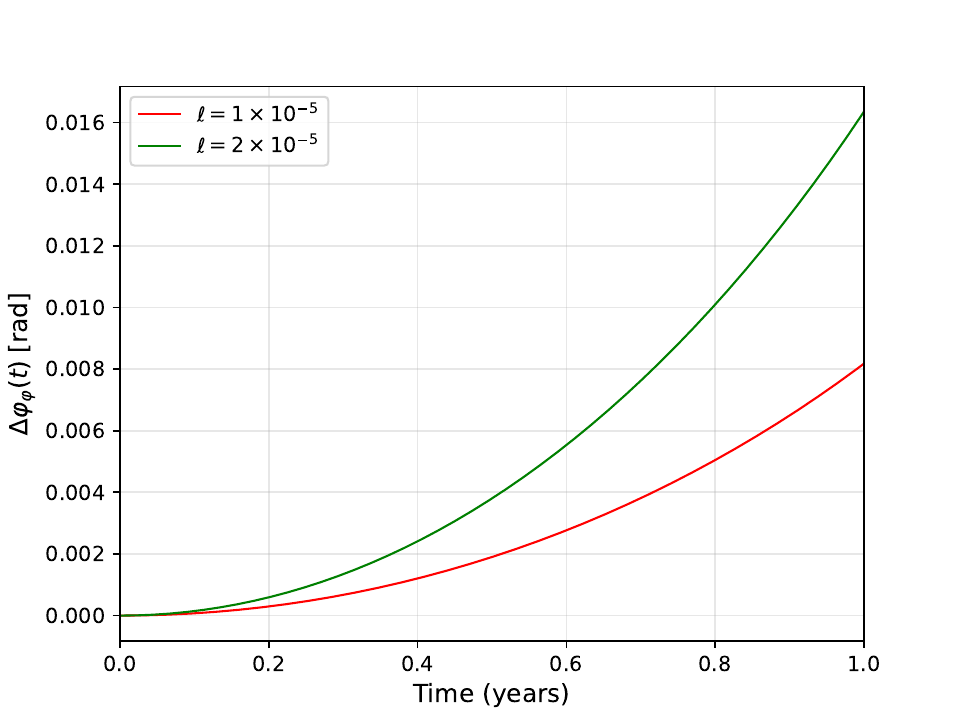}
	\end{minipage}
\caption{
		{Waveform and accumulated phase differences for an EMRI evolved over $T=1~\mathrm{yr}$. \textit{Top left}: the plus polarization $h_{+}(t)$ during the initial interval $t\in[0,10^{-4}]~\mathrm{yr}$. \textit{Top right}: the late-time interval, shown as $(t-t_0)/(10^{-4}\,\mathrm{yr})$ with $t_0=0.9999~\mathrm{yr}$. The waveform curves correspond to $\ell=0$ (blue), $1\times10^{-5}$ (red), and $2\times10^{-5}$ (green). \textit{Bottom left}: $\Delta\varphi_r(t)=\varphi_r^{(\ell)}(t)-\varphi_r^{(0)}(t)$. \textit{Bottom right}: $\Delta\varphi_\varphi(t)=\varphi_\varphi^{(\ell)}(t)-\varphi_\varphi^{(0)}(t)$. The phase differences are shown for $\ell=1\times10^{-5}$ and $2\times10^{-5}$. The remaining parameters are $(M,\mu,p,e)=\left(10^6 M_\odot,\,10 M_\odot,\,10,\,0.2\right)$, where $p$ is dimensionless.}
}
	
	\label{fig:orbit}

\end{figure}
We compare the plus polarization $h_{+}(t)$ of EMRI waveforms for several representative values of $\ell$ in Fig.~\ref{fig:orbit}. {The three waveforms overlap during the initial interval but separate in the late-time interval. The lower panels show the accumulated phase differences $\Delta\varphi_r(t)=\varphi_r^{(\ell)}(t)-\varphi_r^{(0)}(t)$ and $\Delta\varphi_\varphi(t)=\varphi_\varphi^{(\ell)}(t)-\varphi_\varphi^{(0)}(t)$. The magnitude of each phase difference grows during the one-year evolution and is larger for $\ell=2\times10^{-5}$ than for $\ell=1\times10^{-5}$. Although $\Omega_\varphi$ has no explicit $\ell$ dependence at fixed $(p,e)$, the $\ell$-dependent evolution of $p(t)$ and $e(t)$ generates the accumulated difference in $\varphi_\varphi(t)$. The mismatch analysis below quantifies the resulting waveform separation.}

\subsection{Parameter Estimation}
\label{sec:WB}

To connect with observations, it is necessary to convert the plus and cross polarizations of the GW into the LISA detector response.
The response of {LISA} to the incident GW is modeled by two independent Michelson-type channels $\alpha=I,II$, with time-domain strain~\cite{Barack:2003fp}
\begin{eqnarray}
	h_{\alpha}(t)=\frac{\sqrt{3}}{2}\left[F^{+}_{\alpha}(t)h^{}_{+}(t)+F^{\times}_{\alpha}(t)h^{}_{\times}(t)\right]\,,
\end{eqnarray}
where $F^{+,\times}_{\alpha}(t)$ are the antenna pattern functions.

To quantify how distinguishable two EMRI signals are, we compute the
noise--weighted overlap between two LISA responses $h_a(t)$ and
$h_b(t)$. Working in the frequency domain, the corresponding inner
product is defined as
\begin{equation}
	(h_a|h_b)
	= 2 \int
	\frac{h_a^{*}(f)\,h_b(f) + h_a(f)\,h_b^{*}(f)}{S_n(f)}\,{\rm d}f \,,
\end{equation}
where $S_n(f)$ is the one--sided power spectral density of the
detector noise~\cite{LISA:2017pwj}.
Then, the normalized overlap  between the two waveforms is
\begin{equation}	
	\mathcal O(h_a,h_b)
	= \frac{(h_a|h_b)}{\sqrt{(h_a|h_a)\,(h_b|h_b)}} \,,
\end{equation}
and we define the corresponding mismatch as
\begin{equation}
	\mathcal M(h_a,h_b) = 1 - \mathcal O(h_a,h_b) \,.
\end{equation}
{A mismatch of $\mathcal M=0$ indicates that the signal and template are perfectly matched, whereas $\mathcal M=1$ indicates that they are completely mismatched.
}

\begin{figure}[tbp]
	\centering
	\begin{minipage}[b]{0.5\textwidth}
		\centering
		\includegraphics[width=\textwidth]{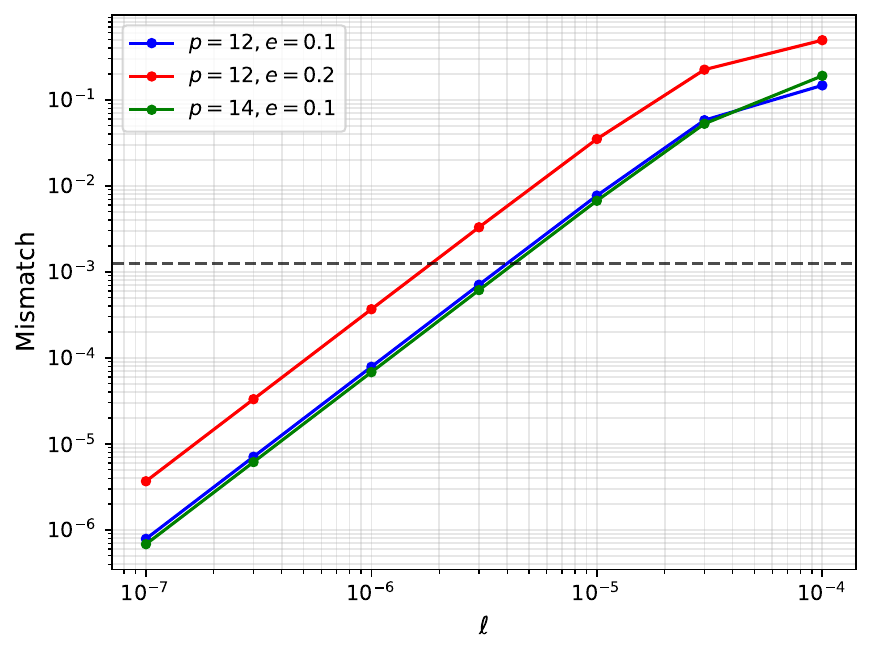}
	\end{minipage}

	\caption{		
		Mismatch as a function of $\ell$ between the reference waveform ($\ell=0$) and the modified waveform ($\ell\neq 0$), shown for three representative initial orbital configurations $(p,e)$ as labeled in the legend. The horizontal dashed line indicates the nominal detectability threshold {$\mathcal M=1.25\times 10^{-3}$~\cite{Kumar:2024our}}; points above this line correspond to waveform pairs that are distinguishable at the adopted criterion. The remaining parameters are chosen to be identical to those in Fig.~\ref{fig:orbit}.
	}
	\label{fig:mismatch}	
\end{figure}

{Figure~\ref{fig:mismatch} shows the mismatch between the reference waveform ($\ell=0$) and the modified waveform ($\ell\neq0$) for three initial configurations $(p,e)$. Over the plotted range, the mismatch increases monotonically with $\ell$. At fixed $p=12$, the $e=0.2$ configuration has a larger mismatch than the $e=0.1$ configuration. At fixed $e=0.1$, the curves for $p=12$ and $p=14$ are close and change ordering over the plotted range. We therefore draw no monotonic conclusion about the dependence on $p$. These trends quantify the sensitivity to $\ell$ within the adopted waveform model.}

\begin{table}[htbp!]
	\centering
	\begin{tabular}{c|c|c|c}
		\hline
		\hline
		{Parameters} & Description&Injected values & {Priors ($\delta=0.01$)} \\ 
		\hline
		\hline
		$\ln (M/M_\odot)$ &Mass of supermassive black hole &$\ln(10^6)$& $\mathcal U[\ln M * (1- \delta), \ln M * (1+ \delta)]$ \\ \hline
		$\ln (\mu /M_\odot)$ & Mass of small black hole& $\ln(10)$& $\mathcal U[\ln \mu* (1- \delta), \ln \mu* (1+ \delta)]$ \\ \hline
		$p_i$ & {Initial dimensionless semi-latus rectum} &$12$& $\mathcal U[p_i * (1- \delta), p_i * (1+ \delta)]$ \\ \hline
		$e_i$ & Initial eccentricity&$0.2$& $\mathcal U[e_i * (1- \delta), e_i * (1+ \delta)]$ \\ \hline
		$\ell$ & bumblebee parameter &0.0025& $\mathcal U[-0.5, 0.5]$ \\ 
		\hline
		\hline
	\end{tabular}
	\caption{Injected values and Prior distributions on the waveform parameters 
		used for MCMC posterior sampling.
		$\mathcal{U}$ denotes the uniform distribution.\label{tab:priors}}
\end{table}

{The waveform examples and the Bayesian injection use different values of $\ell$ for different purposes. Figure~\ref{fig:orbit} uses $\ell=10^{-5}$ and $2\times10^{-5}$ to display the phase separation accumulated over one year. The value $\ell=0.0025$ in Tab.~\ref{tab:priors} is a representative nonzero injection used to test whether an EMRI observation can recover the bumblebee parameter jointly with the intrinsic source parameters. It was not selected because this particular value is theoretically or observationally preferred. A nearby nonzero value would serve the same methodological purpose, although the quantitative posterior can depend on the injection. Both the waveform values and the MCMC injection are illustrative and are not intended to represent observationally allowed values under the universality assumption discussed above. The uniform prior $\mathcal U[-0.5,0.5]$ is likewise a computational choice within the metric domain $\ell>-1$. An analysis designed to forecast a bound around GR would instead use an injection at $\ell=0$.}

{We next infer the posterior distribution of the sampled parameters. EMRI likelihoods can contain strong correlations and multiple local maxima~\cite{Chua:2018yng,Ali:2012zz}. Here we vary the five intrinsic parameters $(\ln M,\ln\mu,p,e,\ell)$, which control the orbital dynamics and secular phase accumulation, while fixing the sky position, orientation, luminosity distance, and initial phases. In our previous full Bayesian EMRI study, reducing a 13-dimensional analysis to the intrinsic-parameter subspace did not qualitatively change the intrinsic posterior shapes or correlations, although the target-parameter upper limit tightened from $1.49\times10^{-3}$ to $1.11\times10^{-3}$, a few-tens-of-percent effect (approximately $26\%$ relative to the full result)~\cite{Xia:2026aty}. Although that study concerned a different physical parameter, the comparison indicates that an intrinsic-parameter-only analysis can provide a useful computationally efficient first estimate. A fully realistic inference should nevertheless also vary the extrinsic and phase parameters, which can broaden or distort the posterior through the detector response, signal-to-noise ratio, and additional parameter correlations.}

 In a Bayesian framework, the information on the source parameters is encoded in the posterior probability distribution, which is given by Bayes' theorem,
\begin{equation}
	p(\Theta|D)=\frac{p(D|\Theta)\,\pi(\Theta)}{p(D)},
\end{equation}
where $\pi(\Theta)$ is the prior distribution, $p(D|\Theta)$ is the likelihood, and $p(D)$ is the evidence. For GW data with stationary Gaussian noise, the likelihood is usually constructed from the noise-weighted inner product between the {injected waveform} and the template waveform.  In practice, the posterior is explored numerically using MCMC sampling, from which we can obtain marginalized distributions and credible intervals for both the standard source parameters and the bumblebee parameter.

The basic procedure is straightforward. For each parameter vector, we generate the corresponding waveform and detector response, and construct the likelihood by using the noise-weighted inner product
\begin{equation}
	\mathcal{L}(D|\Theta)\propto \exp\left[-\frac{1}{2}(D-h(\Theta)\,|\,D-h(\Theta))\right],
\end{equation}
Combined with suitable prior distributions listed in Tab.~\ref{tab:priors}, Bayes' theorem yields the posterior probability density of the model parameters. We then use MCMC sampling to explore this posterior in the five-dimensional parameter space. Once the chains have converged, the resulting samples are used to construct corner plots and marginalized posterior distributions, from which we obtain the credible intervals. {These intervals are rapid, approximate estimates conditional on the stated source configuration, the fixed extrinsic and phase parameters, and the adopted phenomenological waveform.}

\begin{figure}[tbp]
	\centering
	\begin{minipage}[b]{0.5\textwidth}
		\centering
		\includegraphics[width=\textwidth]{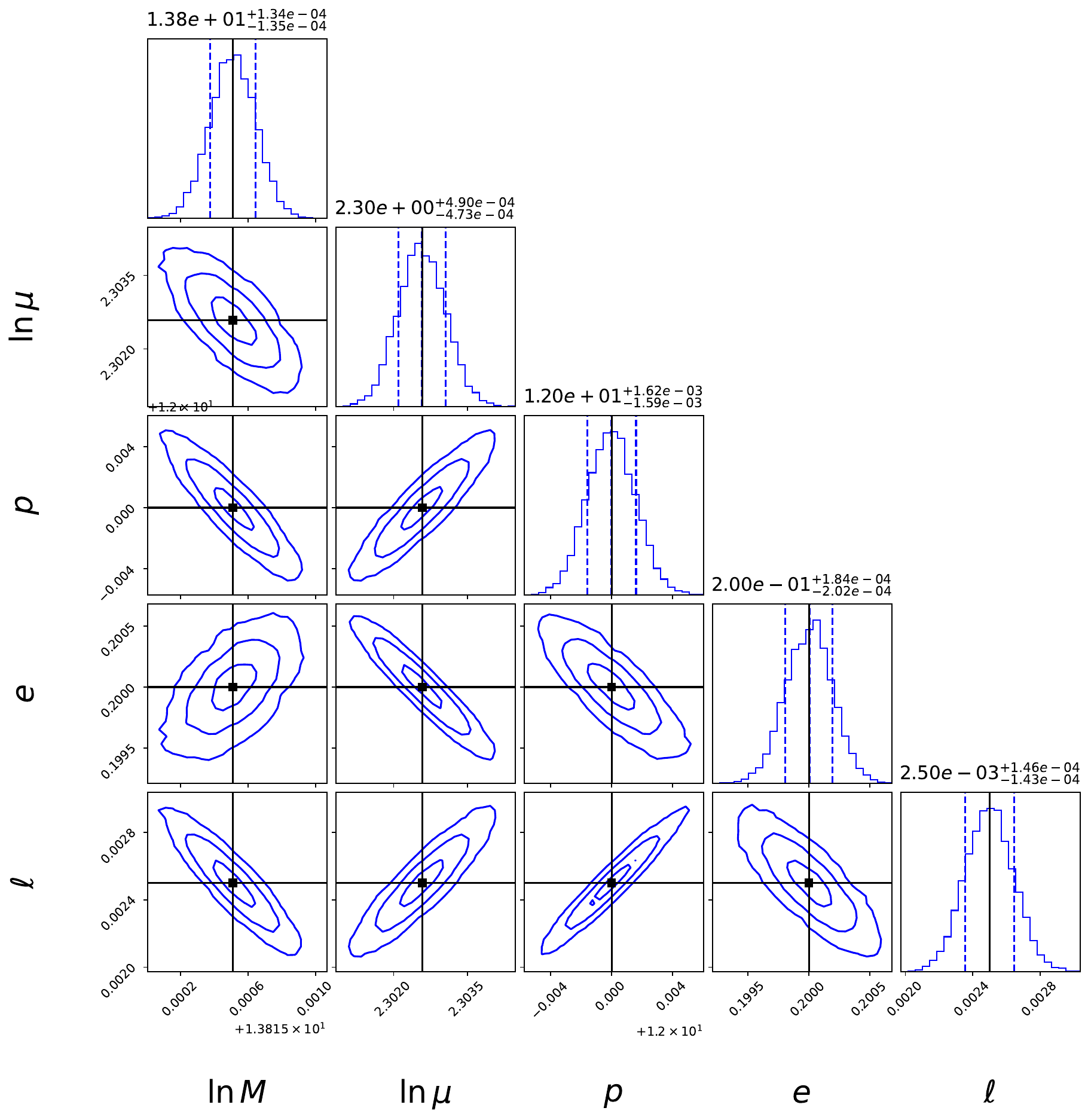}
	\end{minipage}
	\caption{Corner plot of the marginalized posterior distributions for the intrinsic parameter set $(\ln M,\ln \mu,p,e,\ell)$ inferred from an MCMC analysis of a simulated one-year LISA observation. One- and two-dimensional marginalized posteriors are shown on the diagonal and off-diagonal panels, respectively; the contours denote the $68.3\%$, $95.4\%$, and $99.7\%$ joint credible regions. The vertical and horizontal black solid lines mark the injected values. The injected initial conditions and the adopted prior ranges are summarized in Tab.~\ref{tab:priors}}	
	\label{fig:corner}	
\end{figure}

Fig.~\ref{fig:corner} shows the corner plot of the marginalized posterior distributions for the intrinsic parameters $(\ln M,\ln \mu,p,e,\ell)$ obtained from the MCMC analysis. From this figure, we {find} that the intrinsic EMRI parameters can be well determined through the MCMC analysis. The posterior peaks {all} lie close to the injected values, and all the parameters can be recovered within the $1\sigma$ credible interval.
More specifically, the recovered parameters are
$
\ln M = 13.815511\, ^{+1.34\times 10^{-4}} _{-1.35\times 10^{-4}}
$, 
$
\ln \mu = 2.302572\,^{+4.90\times 10^{-4}}_{-4.73\times 10^{-4}}
$,
$
p = 11.99999\,^{+1.62\times 10^{-3}}_{-1.59\times 10^{-3}}
$, 
$
e = 0.200047\,^{+1.84\times 10^{-4}}_{-2.02\times 10^{-4}}
$ and 
$\ell = 0.002499\,^{+1.46\times 10^{-4}}_{-1.43\times 10^{-4}}
$.
{All injected values lie within the reported $1\sigma$ credible intervals. For the illustrative injection considered here, the marginalized posterior of $\ell$ is centered near the injected value and has a statistical width of order $\mathcal{O}(10^{-4})$. This result describes parameter recovery in the reduced five-dimensional analysis; it is not a general LISA bound on bumblebee gravity. Sampling additional parameters may broaden the interval, while a self-consistent radiation model could also shift the posterior.}

\section{Conclusion}
\label{sec:con}

In this work, we have investigated GW signals from eccentric EMRI in a Schwarzschild-like black hole spacetime arising in bumblebee gravity with a bumblebee parameter $\ell$. Starting from the geodesic motion of a test particle in this modified background, we analyzed how the {bumblebee gravity modifies} the conservative orbital dynamics and the 
{corresponding orbital frequencies}. 
{We find that the LSB parameter affects only the radial frequency $\Omega_r$, entering as an overall rescaling factor, while leaving the azimuthal frequency $\Omega_\phi$ unchanged.
} We then constructed the GW energy and angular-momentum fluxes based on the multipolar radiation formula for 
{the eccentric orbit}. 
{We find that the LSB parameter modifies both the energy flux and the angular momentum flux, and that these corrections already enter at the leading order.
} 
{To assess the detectability of bumblebee gravity, these} modifications were subsequently incorporated into the AAK framework to generate EMRI waveforms. 

{For the representative configurations, the bumblebee and Schwarzschild waveforms overlap over the initial interval and develop a phase difference by the late-time interval. Over the range shown in Fig.~\ref{fig:mismatch}, the mismatch increases monotonically with $\ell$. At fixed $p=12$, the $e=0.2$ configuration has a larger mismatch than the $e=0.1$ configuration. At fixed $e=0.1$, the curves for $p=12$ and $p=14$ do not support a uniform directional dependence on $p$.}

{Within the reduced five-dimensional inference, the sampled parameters are recovered near their injected values. For the illustrative nonzero injection, we obtain $\ell = 0.002499\,^{+1.46\times10^{-4}}_{-1.43\times10^{-4}}$. This interval depends on the fixed parameters and the phenomenological waveform; it is not a source-independent upper bound. Comparisons with Solar-System or stellar-orbit constraints require the same parameter $\ell$ to apply across all regimes. Under this assumption, the present result probes a different gravitational regime but does not supersede the Cassini sensitivity reported for the same static metric~\cite{Casana:2017jkc}.}




{The radiation-reaction sector is modeled by evaluating the standard quadrupole formula on the modified orbit. A self-consistent calculation in bumblebee gravity may contain theory-specific flux corrections, bumblebee-field radiation, additional polarizations, propagation effects, and further phase shifts. The analysis also fixes the sky position, orientation, distance, and initial phases. A broader parameter space, a self-consistent radiation model, and an $\ell=0$ injection are needed before this framework can yield an observational upper bound.}

{Within these limitations, the accumulated dephasing found for the representative configurations supports further study of eccentric EMRIs as strong-field probes of bumblebee gravity.}

\appendix
\section{Convergence with respect to the harmonic cutoff}
\label{app:nmax_convergence}

{To test the numerical truncation of the harmonic sum in Eq.~\eqref{eq:h_decomposition}, we repeat the one-year mismatch calculation and posterior analysis using $n_{\max}=6$, $12$, $18$, and $24$. Increasing the cutoff from $n_{\max}=6$ to $12$ changes the mismatch from approximately $0.627566$ to $0.627624$, an absolute difference of $5.8\times10^{-5}$ and a relative change of approximately $9.2\times10^{-5}$. The results for $n_{\max}=12$, $18$, and $24$ agree to the displayed precision. The marginalized $\ell$ posteriors for all four cutoffs nearly overlap and remain centered near the injected value. The small mismatch variation and stable posterior indicate that the reported results are insensitive to the cutoff over the tested range and support the use of $n_{\max}=6$ for the configuration considered here.}

\begin{figure}[!ht]
	\centering
	\begin{minipage}[b]{0.48\textwidth}
		\centering
		\includegraphics[width=\textwidth]{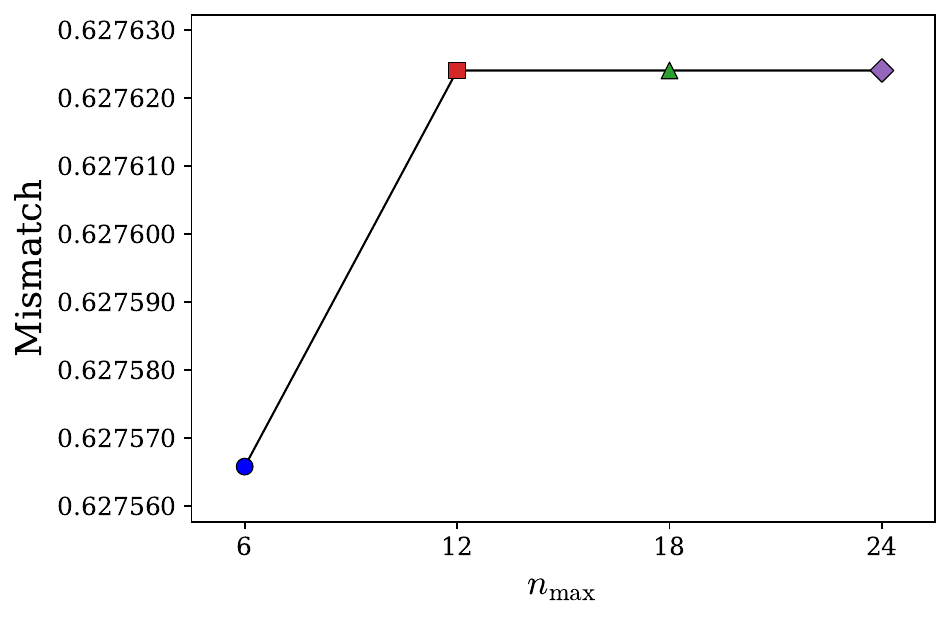}
	\end{minipage}
	\hfill
	\begin{minipage}[b]{0.48\textwidth}
		\centering
		\includegraphics[width=\textwidth]{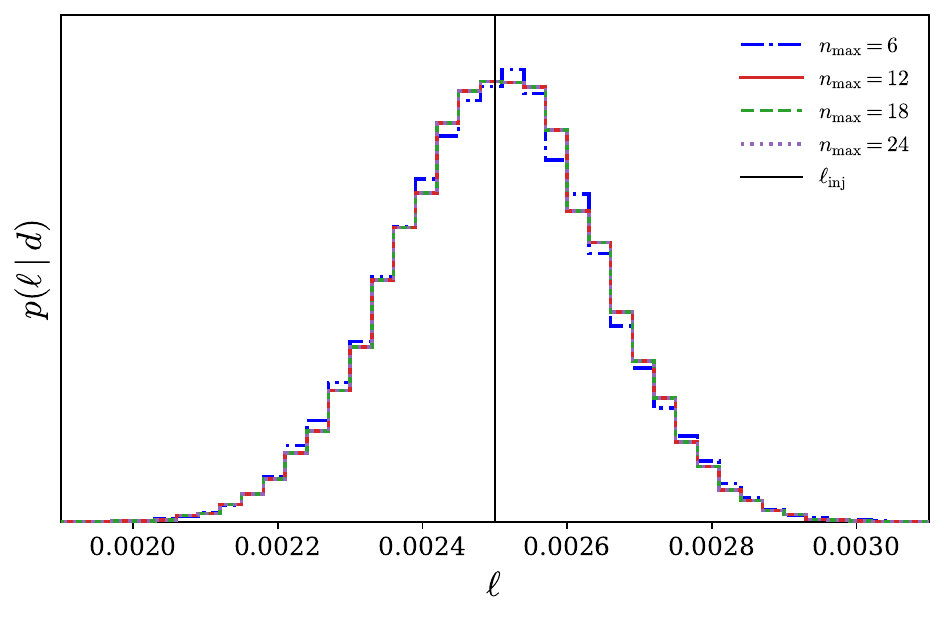}
	\end{minipage}
	\caption{{Convergence with respect to the harmonic cutoff $n_{\max}$. \textit{Left}: mismatch for the one-year waveform at $n_{\max}=6$, $12$, $18$, and $24$. \textit{Right}: marginalized posterior of $\ell$ for the same cutoff values; the vertical black line marks the injected value $\ell=0.0025$. The mismatch varies by less than $6\times10^{-5}$ between $n_{\max}=6$ and the higher tested cutoffs, and the posterior curves nearly overlap.}}
	\label{fig:nmax_convergence}
\end{figure}

\acknowledgments

This work is supported by the National Natural Science Foundation of China (Grants Nos. 12547143, 12275079, 2035005 and 12447156), China Postdoctoral Science Foundation (Grant No. 2025M773339), the National Key Research and Development Program of China (Grant No.
2020YFC2201400) and the innovative research group of Hunan Province (Grant No. 2024JJ1006).

\cleardoublepage
\bibliographystyle{apsrev4-2}
\bibliography{reference}

\end{document}